\documentclass[12pt,twoside]{article}   %
\usepackage[super,sort,comma]{natbib}

\usepackage{fancyhdr}		
\usepackage[section]{placeins}   %

\usepackage{graphicx}
\usepackage{amsmath}
\usepackage{amsfonts}
\makeatletter \renewcommand\@biblabel[1]{$^{#1}$} \makeatother
\newcommand{\note}[1]{\mbox{}\\ \noindent \rule{16cm}{0.5mm} \\
{\em #1} \\ \noindent \rule{16cm}{0.5mm}
\typeout{    }
\typeout{***********note active on this page *************************}
\typeout{Note: #1  }
\typeout{****************************************end Note}
}

\newcommand{\cen}[1]{\begin{center} #1 \end{center}}

\usepackage[mathlines]{lineno}

\usepackage{hyperref}
\hypersetup{ colorlinks,
    citecolor=blue,
    filecolor=blue,
    linkcolor=blue,
    urlcolor=blue
}

\usepackage{xcolor}

\definecolor{gray}{rgb}{0.6,0.6,0.6}
\definecolor{red}{rgb}{0.85,0,0}
\definecolor{green}{rgb}{0,0.85,0}
\definecolor{blue}{rgb}{0,0,0.85}
\definecolor{beige}{rgb}{0.92,0.87,0.78}
\usepackage[all]{hypcap}    

\begin{document}

\cen{\sf {\Large {\bfseries Seven-Probe Fiber Detector for Time-Resolved Source Tracking in HDR-Brachytherapy: Experimental Evaluation} \\  
\vspace*{10mm}
Mathieu Gonod$^1$, Miguel Angel Suarez$^2$, Samir Laskri$^3$, Gwenaël Rolin$^4$, Karine Charriere$^5$, Emmanuel Dordor$^2$, Julien Crouzilles$^3$, Thomas Lihoreau$^6$, Lionel Pazart$^5$, Jean-François Vinchant$^3$, Léone Aubignac$^1$ and Thierry Grosjean*$^2$} \\

$^1$Centre Georges François Leclerc (CGFL) - Dijon, France\\
$^2$Université Marie et Louis Pasteur, CNRS, Intitut FEMTO-ST, F-25000 Besançon, France\\
$^3$SEDI-ATI Fibres Optiques, 8 Rue Jean Mermoz, 91080 Évry-Courcouronnes, France\\
$^4$Université Marie et Louis Pasteur, CHU Besançon, UMR INSERM 1098 RIGHT, INSERM CIC1431, F-25000 Besançon, France\\
$^5$Université Marie et Louis Pasteur, CHU Besançon, UMR INSERM 1322 LINC, INSERM CIC1431, F-25000 Besançon, France\\
$^6$Université Marie et Louis Pasteur, CHU Besançon, Inserm CIC 1431, SINERGIES (UR4662), Tech4Health network (F-CRIN), F-25000 Besançon, France
\vspace{5mm}\\
Version typeset \today\\
}

\pagenumbering{roman}
\setcounter{page}{1}
\pagestyle{plain}
*Author to whom correspondence should be addressed. email: thierry.grosjean@univ-fcomte.fr\\

\begin{abstract}

\noindent {\bf Background:} \textit{In vivo} dosimetry (IVD) is increasingly recognized as a critical tool for verifying treatment delivery in HDR-brachytherapy (HDR-BT). Time-resolved techniques, such as source tracking, enables real-time error detection and reduces uncertainties in dose delivery. An ideal IVD system must accurately track the source across extended volumes along the source’s travel range and measure dwell times as short as fractions of a second.\\

{\bf Purpose:}  This study evaluates a compact biocompatible Seven-probe Scintillator Detector (7SD) for monitoring HDR-BT treatment sequences across a range of dwell times and source-probe spacings representative of most HDR-BT techniques. The capability of the 7SD to detect source insertion errors is also assessed.   \\

{\bf Methods:} The SSD comprises seven detection cells made of Gd$_2$O$_2$S:Tb, each measuring 0.28 $\pm$ 0.02 mm in diameter and 0.43 $\pm$ 0.02 mm in length, coupled to the microstructured tips of silica optical fibers (110-micron diameter). The probes, spaced 15 mm apart along the fiber axis, are organized into a bundle with a total diameter of less than 0.45 mm. The SSD was tested in a $50 \times 50 \times 30$ cm$^3$ phantom using a MicroSelectron 9.1 Ci Ir-192 HDR afterloader connected to a BT stainless steel interstitial needle. Detection signals were acquired at 0.06-second intervals with an sCMOS camera equipped with a chromatic filter to eliminate spurious Cerenkov signals. The probe was experimentally calibrated in 2D without relying on the AAPM-TG43 formalism.  Monitoring of dwell times and positions was performed by combining detection signals from all seven probes. Cytotoxicity tests were performed to confirm the biocompatibility of the 7SD\\

{\bf Results:} A total of 4,040 dwell positions were analyzed, covering source-probe spacings from 10 to 36 mm and a source travel range of 62 mm, with dwell times ranging from 0.1 to 19.5 s. The 7SD successfully identified 99.5\% of the dwell positions. In cylindrical coordinates, the measured dwell positions deviated from the planned values by 0.224 $\pm$ 0.155 mm (radial) and 0.077 $\pm$ 0.181 mm (axial, source travel axis). The average deviation from planned dwell times was 0.006 $\pm$ 0.061 s. 99.4\% of the dwell positions were measured within the 1 mm reliability threshold. The remaining 0.6\% of deviations consistently occurred at the initial dwell position of treatment sequences and appear to stem from a systematic source positioning error by the afterloader. Additionally, the detector accurately identified intentional needle mispositioning scenarios with sub-millimeter accuracy.\\

{\bf Conclusions:} A 7SD based on Gd$_2$O$_2$S:Tb, coupled with an sCMOS camera, demonstrates its suitability for time-resolved IVD in monitoring HDR-BT treatments. The 7SD substantially extends the monitoring volume along both the source’s axial path and the orthogonal radial direction, enabling accurate source tracking across treatment geometries representative of a wide variety of HDR-BT procedures. By detecting subtle afterloader malfunctions, the 7SD offers a precise and valuable means of evaluating the integrity of HDR-BT setups and procedures. Its high-resolution monitoring capabilities, confirmed biocompatibility, scalability, and compatibility with standard BT needles, catheters, and applicators underscore its potential for clinical implementation.

\end{abstract}
\note{This is a sample note.}

\newpage

\pagenumbering{arabic}
\setcounter{page}{1}
\pagestyle{fancy}

\section{Introduction}

High-Dose-Rate brachytherapy (HDR-BT) is a widely used radiotherapy technique that delivers high doses of radiation directly to the tumor while sparing surrounding healthy tissues. By using a temporary radioactive source, such as Iridium-192, positioned precisely within or near the tumor via applicators, catheters or needles, HDR BT achieves exceptional dose localization and treatment efficacy, making it a standard approach for cancers like prostate, cervical, and breast cancer. However, the high precision required in HDR BT also makes it susceptible to potential errors, such as incorrect source positioning, dwell time deviations, or applicator misplacement. These errors can lead to suboptimal dose delivery, resulting in reduced treatment efficacy or unintended exposure to healthy tissues. Given the high dose gradients involved and the proximity to critical organs, even small deviations can have significant clinical consequences.

To address these challenges, \textit{in vivo} dosimetry (IVD) has emerged as a promising tool for real-time verification of treatment delivery \cite{fonseca:piro20,verhaegen:phiro20}. IVD enables direct monitoring of dose delivery during the procedure, providing immediate feedback on the treatment. This capability enhances patient safety by promptly identifying and correcting errors, ensuring optimal therapeutic outcomes. Integrating a scintillating cell with an optical fiber has shown considerable potential for developing compact detectors capable of time-resolved monitoring of dwell times, source positions, and/or delivered doses \cite{lambert:pmb06,lambert:mp07,andersen:mp09,belley:brachy18,johansen:pm19,kertzscher:ro11,jorgensen:mp21,johansen:brachy18,jorgensen:mp21_2,gonod:pmb22}. Moreover, by comparing measured dose values to those predicted by the treatment planning system (TPS), some source insertion errors have been successfully determined \cite{johansen:brachy18}. However, these single-probe detectors are constrained by a limited monitoring volume.

Integrating multiple detection sites into a single BT needle or catheter is essential for expanding the monitoring volume along the source's travel range, while enabling precise real-time treatment delivery monitoring in 2D or 3D. One approach utilizes arrays of independent fiber-based scintillation detectors \cite{wang:rm14,guiral:mp16,cartwright:mp10}. Another strategy integrates up to three scintillation cells into a single optical fiber, enabling multiple detection points within a single BT needle or catheter to monitor dose deposition \cite{therriault:mp11,therriault:mp13}, as well as dwell times and dwell positions \cite{linares:mp20}. These multipoint detectors have demonstrated time-resolved two-dimensional (2D) source tracking through triangulation techniques \cite{linares:mp20}. An evolution of this technology combines a three-point detector \cite{linares:mp20} with a single-probe dosimeter, each integrated into a distinct BT needle, to enable three-dimensional (3D) source position tracking through inter-probe triangulation \cite{linares:mp21}. 

Recent advancements in probe miniaturization have enabled the development of six-parallel-probe detectors compact enough to fit into BT needles or catheters \cite{gonod:mp23}. These six-probe detectors provide a versatile platform for implementing diverse treatment monitoring strategies. Such a design streamlines \textit{in vivo} detection architectures while improving detection accuracy. To date, evaluations of this miniaturized multiprobe approach have primarily focused on optimizing one-dimensional (1D) source tracking along the source displacement axis, with constant dwell times of 10 seconds. To fully realize the potential of multiprobe detectors, a comprehensive evaluation of their multidimensional monitoring capabilities under diverse BT conditions is required. This involves testing performance across a broader range of dwell times, down to 0.1 seconds, and source-probe spacings of up to a few centimeters. Such an analysis is essential to establish their reliability and effectiveness for routine clinical practice. Additionally, the ability of these detectors to identify and quantify source insertion errors, such as axial shifts and angular misalignments, is a key factor that must be addressed to ensure accurate treatment delivery.

This study introduces and evaluates a biocompatible Seven-probe Scintillator Detector (7SD) for monitoring HDR-BT treatment sequences with dwell times ranging from 0.1 to 19.5 seconds. The 7SD substantially increases the monitoring volume along both the axial displacement of the HDR-BT source and the orthogonal radial direction, facilitating precise tracking over a wide variety of treatments. The 7SD was encapsulated in a rigid, biocompatible plastic tube with an outer diameter similar to that of standard prostate HDR-BT needles. With a total diameter of less than 0.45 mm, the 7SD is compact enough to be integrated into most HDR-BT needles and catheters, making it suitable for a wide range of clinical applications. The probe assessment focuses on the time-resolved measurement of the ($r,z$) coordinates of the source for 2D source tracking, dwell time measurement, and the dwell position identification rate as a function of dwell time. The reliability level of the 7SD in monitoring dwell position was also studied. Additionnaly, the detector’s ability to detect source insertion errors was evaluated under conditions of needle mispositioning (axial shifts along the (0z) axis) and misalignment (source needle tilt relative to the detector). 

\section{Materials and Methods}

\subsection{The 7-probe detector}

\subsubsection{7SD design and structure}

The 7SD consists of seven independent inorganic scintillating microcells grafted at the tips of seven biocompatible optical fibers forming a 10-meter-long hexagonal-lattice bundle. Each fiber has an outer diameter of 110 microns (with a 100-micron core) and is coated with an 8-micron-thick polyimide protective layer. The combined fiber bundle has a total diameter of 375 microns. Gd$_2$O$_2$S:Tb was selected as the scintillator due to its high luminescence efficiency, stability, linearity, and fast temporal response \cite{qin:ox16,hu:saa18,alharbi:ieee18}. Additionally, Gd$_2$O$_2$S:Tb exhibits low sensitivity to temperature variations in the range of 15$^{\circ}$C to 40$^{\circ}$C \cite{oreilly:ieee20,mclaughlin:rm23}.

To enhance the transfer of luminescence from the scintillators to the guided modes of the optical fibers, each fiber tip was tapered into a controlled shape, serving as an optical analog of a leaky-wave antenna \cite{suarez:ox19}. Details of the probe fabrication process have been described in previous works \cite{gonod:pmb21,gonod:pmb22}. The lengths of the fibers within the bundle were adjusted to maintain a uniform spacing of 15.0 $\pm$ 0.1 mm between successive scintillating cells (Fig. \ref{fig:exp1} (a)). The probes are evenly distributed over a 90 mm span. Each scintillation cell has a diameter of 0.28 $\pm$ 0.02 mm and a length of 0.43 $\pm$ 0.02 mm, resulting in a detection volume of less than 0.02 mm$^3$ (Fig. \ref{fig:exp1} (b)). The diameter of the 7SD structure locally increases to approximately 0.41 mm at the position of the scintillating cells.  

\subsubsection{Encapsulation} \label{sec:encaps}

\begin{figure}[htbp]
\begin{center}
\includegraphics[width=0.5\linewidth]{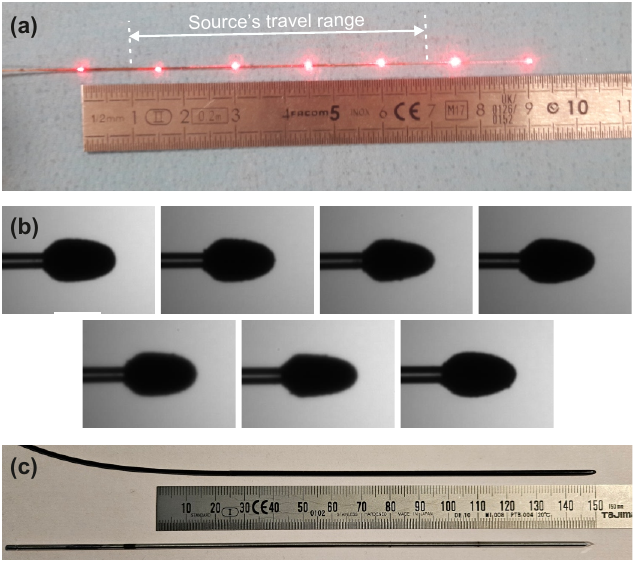}
\caption{The 7SD. (a) Photograph of the 7SD, with the source travel range considered in this study indicated. (b) Microscope images showing the seven tip-integrated scintillation micro-cells composing the 7SD. (c) Photograph of the encapsulated 7SD alongside a commercial HDR-BT interstitial metallic needle for size comparison.}\label{fig:exp1}
\end{center}
\end{figure}

The 7SD is encapsulated in a biocompatible soft plastic tube that provides both optical and mechanical shielding. The final 12.5 cm of the encapsulation consists of a biocompatible rigid tube that securely houses the seven scintillation cells. The outer diameter of the  encapsulated probe, measuring 1.65 mm, is closely matched to the 1.9-mm-outer-diameter stainless steel interstitial needles (Fig. \ref{fig:exp1} (c)). It is well-suited for insertion through a commercial prostate HDR-BT template. This compact design allows the 7SD to be seamlessly integrated into standard clinical HDR-BT procedures, ensuring compatibility with commercial HDR-BT catheters,  needles and applicators.\\

\subsubsection{Cytotoxicity assessment}


Cytotoxicity assessment was performed according to ISO 10993-5 guidelines using the extract method and the MTT assay. To this aim, L929 murine fibroblasts were exposed for 24 h to conditioned media from either an intact or a degraded probe. Briefly, probes were sterilized with ethanol and dry before immersion in culture medium (1 ml, 37°C, 24h). For the degraded condition, mechanical breakage was applied to the probe before immersion. Then, L929 cells were cultured in the presence of conditioned media during 24h.  Cell viability was quantified by measuring optical density after MTT reduction and expressed as a percentage of untreated controls, with a cytotoxic reference (DMSO 10\%). A viability below 70\% was considered as cytotoxic.

\begin{figure}[htbp]
\begin{center}
\includegraphics[width=0.8\linewidth]{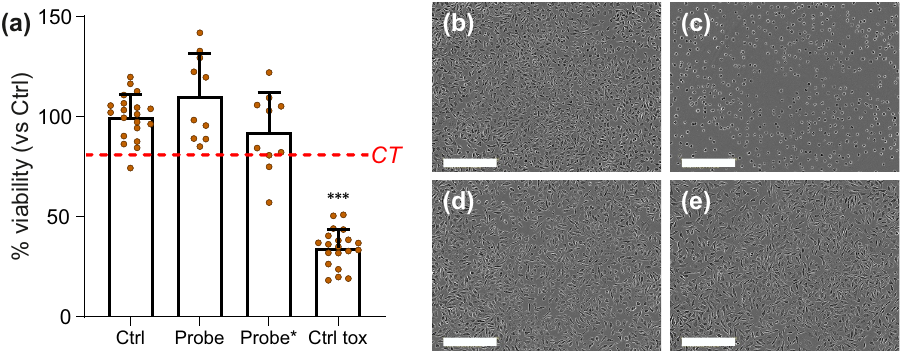}
\caption{Evaluation of the cytotoxicity potential of the intact or degraded probe. (a) Cell viability, measured by MTT assay after 24 hours of culture, was compared to negative (Ctrl) and positive (Ctrl tox) controls ($n=5$, $N=2$). Statistics: one-way ANOVA with Dunnett's test (*** $p<0.001$). CT: threshold of 70\% for cytotoxicity. (d) and (e) Cell morphology after 24 hours of culture in conditioned medium, compared to negative (b) and positive (c) controls: (d) Intact probe, (e) degraded probe. 
Images were captured using the Incucyte S3 microscope (x10). Scalebar: 0.4 mm}\label{fig:cyto}
\end{center}
\end{figure}

Results showed that cell viability remained above 70\% when cells are cultured in the presence of conditioned media from intact or degraded probe (Fig. \ref{fig:cyto}(a)). Such results indicate that no significant cytotoxicity were detected. Morphological analysis confirmed the absence of structural alterations or vacuolization in treated cells compared to the negative control, while the positive control exhibited clear signs of cytotoxicity (see Fig. \ref{fig:cyto}(b-e)).

Given that the materials used for encapsulating the probe and the adhesive are already established as biocompatible, and considering the challenges faced in degrading the outer capsule to assess the cytotoxicity of the Gd$_2$O$_2$S:Tb scintillator, further testing is not warranted. This process provided a preliminary evaluation of the biocompatibility of the 7SD, confirming its safety for potential clinical use.

\subsubsection{Optical readout system}\label{sec:readout}

The encapsulated 7SD was connected to a simple yet effective optical readout system, consisting of an sCMOS camera (Andor Technology, Zyla 4.2 model), a camera objective (Fujinon HF35SA), and a chromatic bandpass filter (Semrock) placed in front of the camera. This setup images the output face of the 7SD while simultaneously filtering out spurious Cerenkov signals (stem effect) \cite{gonod:pmb21,gonod:pmb22,gonod:mp23}. The system operates at a sampling rate $\tau$ of 0.06 seconds, as described in previous work \cite{gonod:mp23}.  

To facilitate real-time acquisition, a custom LabVIEW-based software platform was developed to control the camera. The software automatically defines 16-pixel-diameter Regions of Interest (ROIs) that enclose the seven hexagonally-arranged light spots observed in the images. Image pixels within each ROI were integrated to generate seven independent detection signals, one per probe, sampled at 0.06-second intervals ($\tau$). This automated process enables continuous tracking of scintillation signals during HDR-BT treatment.

\subsection{The measurement setup}

\begin{figure}[htbp]
\begin{center}
\includegraphics[width=0.9\linewidth]{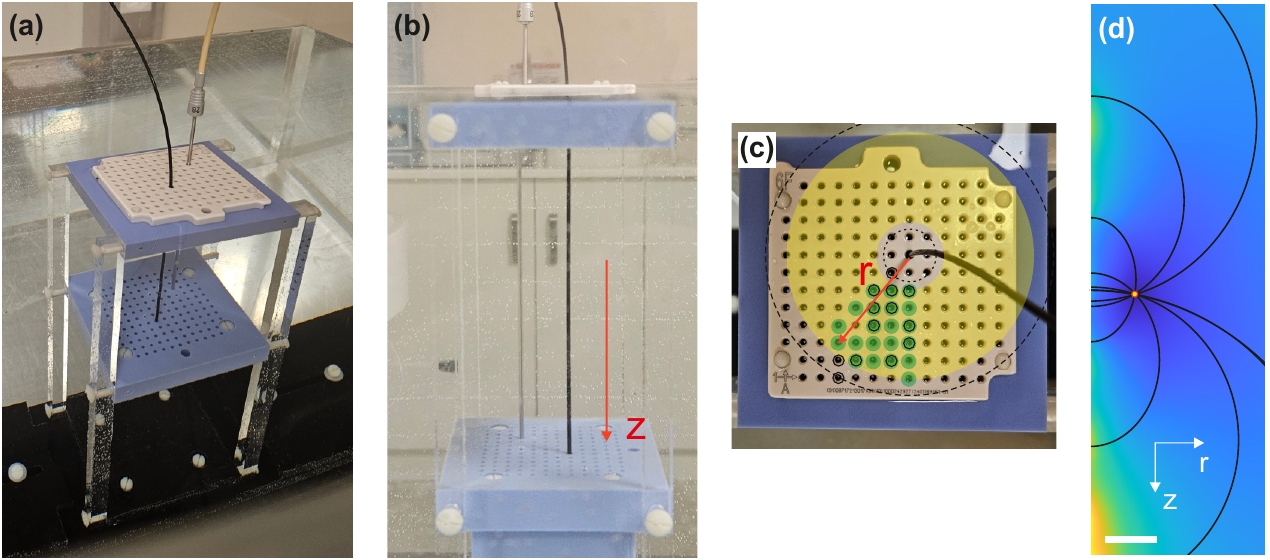}
\caption{(a) Photograph of the measurement setup inside the water tank. (b) Side view of the inter-template region, showing that the source needle and the 7SD are parallel and vertically aligned along the z-axis. (c) Photograph of the top template, where the source insertion holes used for the 7SD calibration and treatment deliveries are marked with black circles and green disks, respectively. The $r$ coordinate is also superimposed on the image. (d) Illustration of the triangulation process used for 2D source positioning at an instant $k \tau$ ($k\in\mathbb{N}$). The seven black elliptical arcs represent the minima of the $\Delta V_i(r, z, k \tau)$ function (Eq. \ref{eq:DV}) for each of the seven probes of the 7SD. The false-color map in the background corresponds to the summation of the seven $\Delta V_i(r, z, k \tau)$ functions, producing a single minimum point at the intersection of the seven positional ellipses. This minimum, marked by a yellow point, determines the source’s location in the ($r,z$) plane.}\label{fig:exp2}
\end{center}
\end{figure}

Treatment monitoring was conducted in a 50 x 50 x 30 cm$^3$ water tank. Irradiation was performed using a MicroSelectron Ir-192 HDR afterloader (9.1 Ci, air kerma strength: 37176 U) connected to a 20 cm stainless steel interstitial needle from Elekta. The needle was positioned in the water tank using a custom-built organic holder consisting of two solid-water needle insertion templates, spaced 13.5 cm apart. Each template features a 12 × 13 array of 5-mm-spaced holes, patterned according to the grid layout of a commercial prostate stepper template from Elekta. To facilitate the identification of source and detector insertion points, the front plate of the commercial prostate stepper template was mounted on top of the upper solid-water template. The plate’s labeled coordinate system, using letters and numbers, provided a standard reference for needle positioning and alignment. To better approach clinical HDR-BT treatment conditions, the front plate was positioned to extend above the water surface. The bottom side of the lower solid-water template was placed in contact with an unprocessed solid-water plate, ensuring that the tips of the source needle and encapsulated probe were aligned at the same height. Images of the holder are presented in Figs. \ref{fig:exp2} (a)-(c). During calibration and treatment measurements, the 7SD is positioned in the insertion hole H8 (see Fig. \ref{fig:exp2}(c)).

During treatment delivery, the source is initially positioned at the deepest location within the water tank, near the bottom solid-water template, and is then retracted toward the water surface in 2.5 mm steps. 

\subsection{System calibration} \label{sec:calib}

Optical signals from the scintillators were recorded simultaneously and processed post-irradiation. The signals were converted into 2D source coordinates using calibration plots. Detector calibration was carried out by measuring signals from the 7SD with the source needle positioned in the 13 insertion holes highlighted with a black circle in Fig. \ref{fig:exp2} (c). For each needle insertion, a series of 41 dwell positions was executed over a distance of more than 10 cm along the vertical (0z) axis (defined in Fig. \ref{fig:exp2}(b)), with a constant inter-dwell spacing of 2.5 mm and a dwell time of 3.7 seconds. Assuming axis symmetry in the environment surrounding the detector, a 2D array of signals corresponding to ($r,z$) positions was defined for each of the seven probes of the 7SD by merging these 13 acquisition sequences. The z-axis (see Fig. \ref{fig:exp2}(b)) was gridded according to the dwell step size, while the r-axis (cf. Fig. \ref{fig:exp2}(c)) followed the grid layout of the source insertion templates. To obtain smooth and continuous calibration plots, the measured data were interpolated using the "interp" function of Matlab software, yielding a sampling rate of 0.1 mm. The resulting seven calibration plots span 10 cm along the z-axis and extend up to 40 mm from the detector along the r-axis. Owing to axial symmetry, the detector calibration encompasses the region outlined by the two dashed lines in Fig. \ref{fig:exp2}(c).

\subsection{Treatment plan}

The irradiation protocol used to evaluate the 7SD is summarized in Table \ref{tab:specif}. This protocol combines five distinct prostate HDR-BT treatment plans, each comprising 10 to 16 treatment sequences conducted in separate implanted needles (see the third column of Table \ref{tab:specif}). Unlike standard prostate HDR-BT, all treatment sequences within a single plan are executed within the same needle. Each treatment plan is repeated four or five times, with the source needle positioned at different distances ($r_{TPS}$) from the detector within the plane of the template (cf. Fig. \ref{fig:exp2}(c)). For each treatment plan, the values of $r_{TPS}$ are provided in the second column of Table \ref{tab:specif}. The 21 source needle positions, combined across all five treatment plans, are marked with green disks in Fig. \ref{fig:exp2}(c). The total number of dwell positions for each plan is listed in the last column of Table \ref{tab:specif}, resulting in a cumulative total of 4040 dwell positions. By symmetry, the evaluation area, highlighted in yellow in Fig. \ref{fig:exp2}(c), extends 36 mm in all directions around the detector.

\begin{table}
\begin{center}
\begin{tabular}{|c|c c c c c|c|c|}
\hline
TP & \multicolumn{5}{c|}{$r_{TPS}$ (mm)} & Number of sequences & Total number of dwells \\
\hline
\hline
1 & 15.8 & 20.6 & 28.3 & 31.6 &  - &  16 & 896 \\

2 & 14.1 & 18 & 26.9 & 30 &  - &  13 & 676\\
3 & 11.8 & 22.4 & 25.5 & 29.2 &  - &  15 & 960 \\
4 & 15 & 20 & 30.4 & 33.5 & - &  15 & 828  \\
5 & 10 & 21.2 & 25 & 32 & 36 & 10 & 680\\
\hline
\multicolumn{6}{|c|}{Total} & \textbf{69} & \textbf{4040}\\
\hline
\end{tabular}
\end{center}
\caption{Irradiation protocol used to evaluate the performance of the 7SD. TP: Treatment plan. }\label{tab:specif}
\end{table}

\begin{figure}[htbp]
\begin{center}
\includegraphics[width=0.65\linewidth]{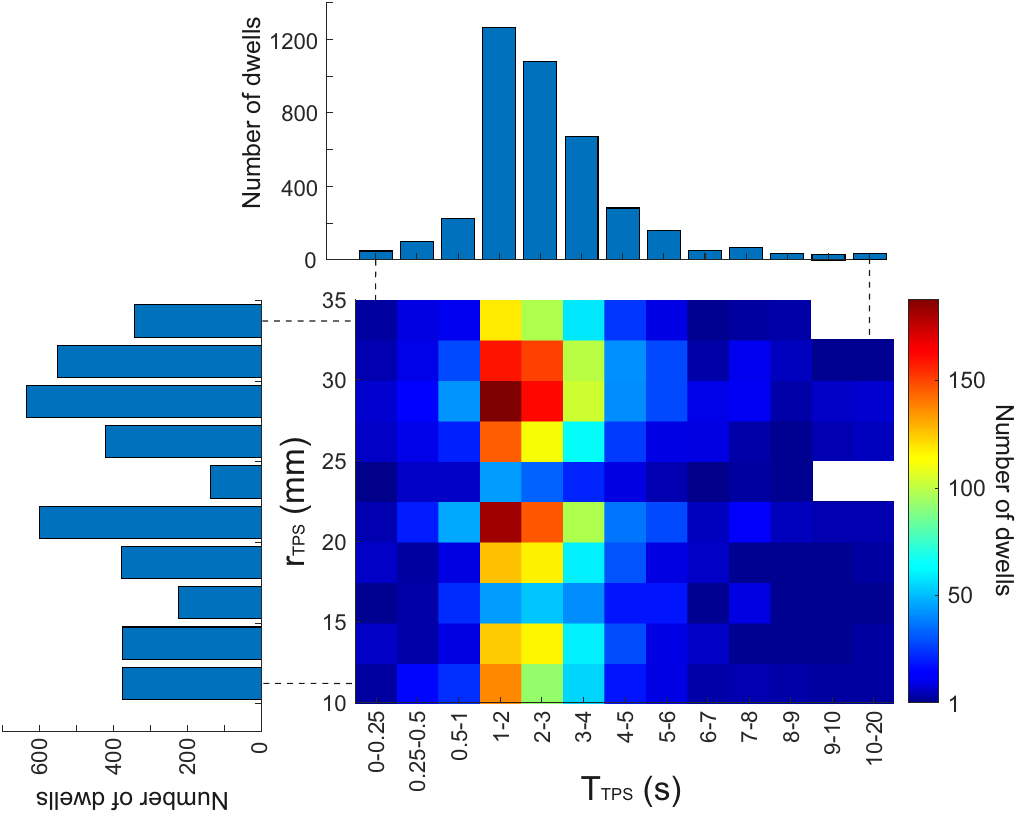}
\caption{Distribution of dwell positions as a function of the planned source-probe spacing ($r_{TPS}$) and planned dwell time ($T_{TPS}$).}\label{fig:distrib}
\end{center}
\end{figure}

\subsection{2D source position tracking using the 7SD system}\label{sec:algo}

During treatment delivery, each probe $P_i$ ($i=[1,7]$) of the 7SD system generates a staircased temporal electric signal $V_i(t)$ sampled at fixed time intervals $\tau=0.06$ seconds. At each time instant $t= k \tau $, where $k \in \mathbb{N}$, the 2D coordinates ($r_s,z_s$) of the source position are determined using the following equation:  

\begin{equation}
    f(r,z,k\tau)=\sum_{i=1}^7 \Delta V_i(r,z,k\tau),\label{eq:f}
\end{equation}

where:
\begin{equation}
    \Delta V_i(r,z,k\tau)=\frac{\left| C_i(r,z) - \kappa V_i(k\tau) \right|}{max(C_i(r,z))}. \label{eq:DV}
\end{equation}

Here, $\kappa$ is a normalization coefficient that compensates for difference in source activity between the calibration and treatment phases. The function $\Delta V_i(r,z,k\tau)$ represents the normalized deviation of the measured signal $V_i(k \tau)$ from the renormalized calibration map $C_i(r, z)$ for each probe. The 2D source position $(r_s, z_s)$ at time $k \tau$ is obtained by identifying the minimum of the composite function $f(r, z, k \tau)$ (Eq. \ref{eq:f}). Each probe $P_i$ corresponds to an individual function $\Delta V_i(r, z, k \tau)$ (Eq. \ref{eq:DV})  whose minimum, ideally zero, defines an elliptical isoprobability contour indicating potential source locations at time $k \tau$. By applying a triangulation method, the sum of the seven $\Delta V_i(r, z, k \tau)$ functions  yields a single minimum point corresponding to the intersection of these elliptical contours, thereby providing the source's position. This triangulation procedure is illustrated in Fig. \ref{fig:exp2}(d).

During treatment, the source moves between successive dwell positions within a time frame of a few tens of milliseconds \cite{fonseca:mp15}. This movement introduces transient phases in the staircase-shaped readout signals, which typically last for one or two acquisition points. To avoid these transient signals in the dwell position analysis, the first and last acquisition points at each dwell position are systematically excluded from the source tracking process.

\subsection{Dwell time measurement}

The dwell time corresponds to the interval between two successive transient phases in the temporal readout signals. It is determined using an edge detection algorithm applied to each probe's readout signal, $V_i(k\tau)$ \cite{canny:ieee86}. The algorithm involves independently convolving each signal $V_i(k\tau)$ with a filter characterized by an impulse response composed of two rectangular functions with opposite signs \cite{canny:ieee86}. This yields seven convolution functions, each displaying a series of narrow peaks. These functions are then summed to produce a combined signal, from which dwell times are extracted as the time intervals between successive peaks. By integrating the signals from all seven probes, the method improves edge detection precision and robustness, reducing the impact of noise from individual probes.

\subsection{Reliability analysis}

The reliability of the 7SD in tracking a HDR-BT source was evaluated by analyzing the rate of detected dwell positions whose overal offset from the planned dwell positions does not exceed 1 mm.  This confidence criterion aligns with the positional accuracy tolerances for the source and X-ray marker, as well as the length tolerances of applicators or treatment tubes, in both HDR and PDR brachytherapy. \cite{dempsey:recomm13}. 

\subsubsection{Systematic reliability analysis}

A systematic reliability analysis of the system was performed by recording signals from the 7SD as the source was incrementally moved along the z-axis within the treatment region.  The irradiation procedure consisted of 12 dwell positions spaced 2.5 mm apart, with a dwell time of 10 seconds at each position. This process was repeated for 12 different source-probe spacings, $r$, ranging from 10 mm to 36 mm.  Using the algorithm detailed in Section \ref{sec:algo}, the distance $R=\sqrt{(r_{TPS}-r_s)^2+(z_{TPS}-z_s)^2}$ between the planned and measured instantaneous source positions was computed as a function of time for the 12 source-probe spacing values. This process yielded 12 time traces of the distance $R$ to the planned source positions, sampled at 0.06-second intervals. Each of these time traces was subsequently resampled 30 times over fixed time intervals, which incrementally increased from 0.1 seconds to 3 seconds. As a result, for each source-probe spacing  $r$, 30 distributions of source-probe position deviations, $\bar{R}$, integrated over time intervals $\Delta t$ ranging from 0.1 to 3 seconds, were generated. For each combination of source-probe spacing $r$ and integration interval $\Delta t$, the fraction of $\bar{R}$-values that fell within the reliability criterion of 1 mm was analyzed.

\subsubsection{Treatment plan analysis}

The overall offset, $R_{dwell}$, between the measured and planned dwell positions, ($\bar{r}_{s},\bar{z}_{s}$) and ($r_{TPS},z_{TPS}$) respectively, is computed as $R_{dwell}=\sqrt{(r_{TPS}-\bar{r}_s)^2+(z_{TPS}-\bar{z}_s)^2}$.

\subsection{Detection of treatment needle mispositioning}

To evaluate the ability of the 7SD to detect errors in source needle positioning, two types of deliberate mispositioning scenarios were introduced: a shift of the source needle along its insertion axis $z$ and a tilt relative to the $z$-axis. In the first scenario, the source shift  was manually displaced by 1, 2, or 3 mm along the $z$-axis. Measurements were performed at a source-probe spacing $r = 22.4$ mm. In the second scenario, the source needle was misaligned by inserting it into two different neighboring holes in the upper and lower solid-water needle insertion templates. 

For both the shift and tilt error analyses, a sequence of 26 dwell positions was investigated, each with a dwell time of 3.7 seconds and spaced 2.5 mm apart. This configuration spanned a 62 mm segment along the $z$-axis, matching the treatment region examined in the other characterizations.

\section{Results}

\subsection{Dwell identification}

Overall, 99.5\% of the dwell positions were successfully identified (see Table \ref{tab:specif_dt}). Notably, all source positions with dwell times of 0.2 seconds or longer were accurately identified, except for one position that was missed due to a time lag in the detection system \cite{gonod:pmb22}. For dwell times of 0.1 seconds, the identification rate was approximately 21\%. 

\begin{table}
\begin{center}
\begin{tabular}{|c|c|c|c|}
\hline
Dwell time & Number of dwells & Number of errors & Identification rate (\%) \\
\hline
\hline
0.1 & 24 & 19 & 20.8 \\
0.2 & 24 & 0 & 100 \\
0.3 & 24 & 0 & 100 \\
0.4 & 47 & 1 & 97.9 \\
$>0.4$ & 3921 & 0 & 100 \\
\hline
\textbf{Total} & \textbf{4040} & \textbf{20} & \textbf{99.5}\\
\hline
\end{tabular}
\end{center}
\caption{Analysis of the 7SD dwell identification as a function of dwell time values.}\label{tab:specif_dt}
\end{table}

\subsection{Dwell position verification}

\begin{figure}[htbp]
\begin{center}
\includegraphics[width=0.95\linewidth]{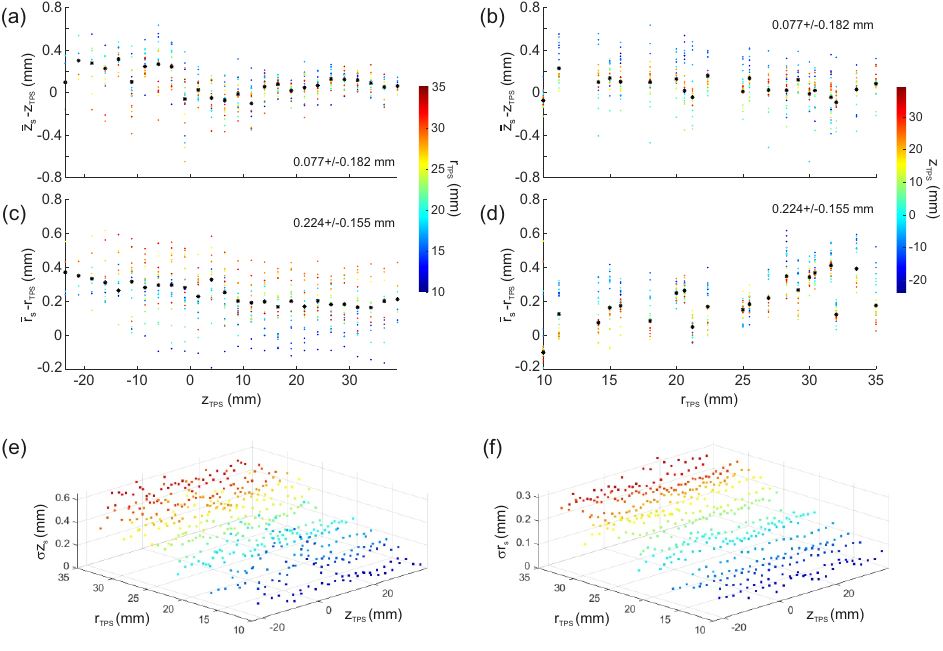}
\caption{(a,b) and (c,d) Deviations $\bar{z}_{s}-z_{TPS}$ and $\bar{r}_{s}-r_{TPS}$ between the measured and planned dwell positions, respectively, as a function of the planned dwell coordinates. Panels (a) and (c) display deviations versus $z_{TPS}$ with $r_{TPS}$ color-coded, while panels (b) and (d) plot deviations versus $r_{TPS}$ with $z_{TPS}$ color-coded. Black markers represent the average deviation for each coordinate value. (e,f) Uncertainty of the 7SD in determining the instantaneous source position, assessed across the entire treatment plan.}\label{fig:dwell_pos_xz}
\end{center}
\end{figure}

Deviations between the planned dwell positions, $(r_{TPS}, z_{TPS})$, and the measured dwell positions, $(\bar{r}_s, \bar{z}_s)$, were assessed separately for each coordinate.  Each point in Figs. \ref{fig:dwell_pos_xz}(a,b) and (c,d) represents the average deviation calculated from all measured dwell positions at the same $(r_{TPS},z_{TPS})$ coordinates. Overall, the measured dwell position coordinates $\bar{r}_s$ and $\bar{z}_s$ deviated from the planned values by 0.224 $\pm$ 0.155 mm  and 0.077 $\pm$ 0.181mm, respectively. 

The uncertainty shown in Figs. \ref{fig:dwell_pos_xz}(e,f) was quantified as the standard deviation (SD) of the distribution of the instantaneous source position coordinates ($r_{s}$,$z_{s}$), detected at a sampling rate of 0.06 s. This analysis accounted for the cumulative dwell times at each spatial coordinate corresponding to repeated dwell positions, resulting from multiple treatment sequences conducted at the same needle locations. The uncertainty of the 7SD to locate the instant source position was less than 0.3 mm along the $r$-axis and 0.5 mm along the $z$-axis, respectively.

\subsection{Dwell time verification}

\begin{figure}[htbp]
\begin{center}
\includegraphics[width=0.95\linewidth]{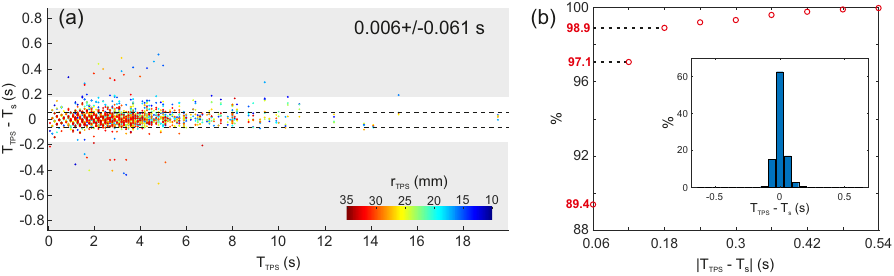}
\caption{(a) Offset between the measured dwell times $T_{s}$ and the corresponding planned dwell times $T_{TPS}$, as a function of $T_{TPS}$. Each data point is color-coded according to the source-probe spacing $r_{TPS}$ (see colorbar). (b) Cumulative percentage of dwell time offsets falling within successive harmonics of the camera’s integration time.}\label{fig:dwell_time}
\end{center}
\end{figure}

Figure \ref{fig:dwell_time}(a) presents the offset between the measured and planned dwell times plotted versus the planned dwell time $T_{TPS}$. About fifty data points, marked within the shaded regions, exceed three acquisition times of the camera. These deviations are attributed to time lag issues in the acquisition system, resulting from the camera and data storage being managed by the same computer under Windows environment\cite{gonod:pmb22}. When all identified dwell positions are considered, including erroneous points, the mean deviation from the planned dwell times is calculated as $0.006 \pm 0.061$ s. A total of 89.4\%, 97.1\% and 98.9\% of the data points fall within one, two, and three integration times of the camera, respectively (see Fig. \ref{fig:dwell_time}(b)). 

\subsection{Reliability evaluation}

\subsubsection{Systematic analysis} 

\begin{figure}[htbp]
\begin{center}
\includegraphics[width=0.9\linewidth]{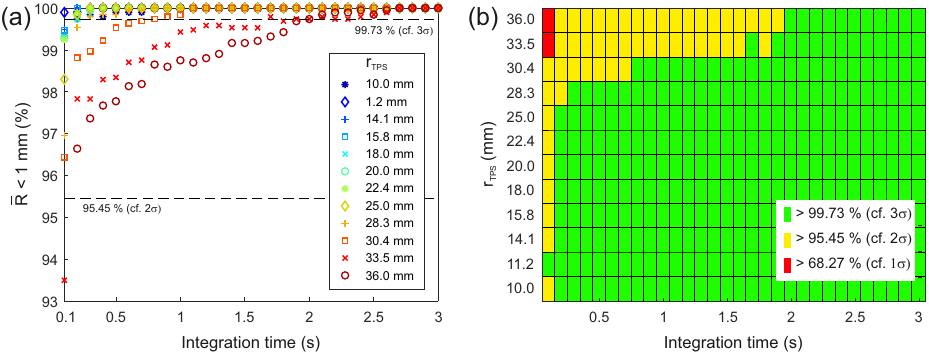}
\caption{(a) Fraction of average source position offsets, $\Bar{R}$, that fall below the defined reliability criterion of 1 mm. These fractions were computed for integration times ranging from 0.1 to 3 seconds, in 0.1-second increments, and for planned source-probe spacings ($r_{TPS}$) spanning from 10 mm to 36 mm. The reliability thresholds of 99.73\% and 95.45\%  are superimposed on the plot for reference. These criteria correspond to the 3$\sigma$ and 2$\sigma$ confidence intervals of a standard normal distribution, respectively. (b) reliability level of the 7SD as a function of the planned source-probe spacing, $r_{TPS}$, and the integration time used to calculate $\Bar{R}$. The reliability is color-coded into three levels, indicating the fractions of average position offsets $\bar{R}$ below 1 mm that are larger than 99.3\%, 95.45\%, and 68.27\% (see legend).}\label{fig:reliability}
\end{center}
\end{figure}

Figure \ref{fig:reliability} shows that for dwell times exceeding 2 seconds, the 7SD maintains a source position accuracy better than 1 mm across source-probe spacings from 10 to 36 mm. For dwell times between 0.9 and 2 seconds, the 7SD ensures this accuracy within a spacing range of 10 to 30.4 mm. Similarly, for dwell times between 0.3 and 0.9 seconds, and for dwell times below 0.2 seconds, the maximum reliability threshold is maintained for source-probe spacings of 10 to 28.3 mm and 10 to 25 mm, respectively.

\subsubsection{Treatment plan}

\begin{figure}[htbp]
\begin{center}
\includegraphics[width=0.7\linewidth]{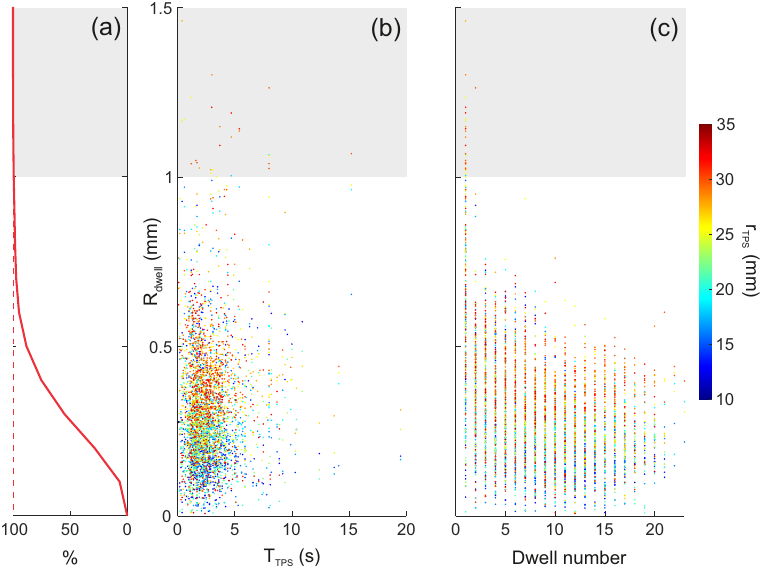}
\caption{(a) Cumulative distribution of measured dwell positions as a function of their overall deviations $R_{dwell}$ from the treatment plan. (b) Overall offsets $R_{dwell}$ of the 4040 measured dwell positions relative to their planned positions, plotted versus the corresponding planned dwell times ($T_{TPS}$) and source-probe spacings ($r_{TPS}$). $r_{TPS}$ is color-coded (see colorbar in (c)). (c) Overall offsets $R_{dwell}$ as a function of the dwell numbers within the treatment sequences and the planned source-probe spacing, $r_{TPS}$, highlighting the relationship between sequential dwell positions and their deviations. Dwell $\#1$ is the last dwell of each treatment sequence. }\label{fig:rel_treat}
\end{center}
\end{figure}

 Figure \ref{fig:rel_treat} presents an analysis of the offsets $R_{dwell}$ between the measured and planned dwell positions across the entire treatment plan. The shaded areas highlight dwell positions which deviate by more than 1 mm from the planned locations. Figure \ref{fig:rel_treat}(a) shows that 88.3\%, 96.7\% and 99.4\% of the measured dwell positions exhibit deviations $R_{dwell}$ below 0.5 mm, 0.75 mm and 1 mm, respectively. Figure \ref{fig:rel_treat}(b) reveals a subtle color gradient in the data points, which is also observed in Figure \ref{fig:rel_treat}(c). Additionally, Fig.\ref{fig:rel_treat}(c) shows that approximately 96\% of deviations exceeding 1 mm occur at the last dwell position of the treatment sequences (dwell $\#1$). This proportion remains at 91\% for deviations larger than 0.77 mm.

\subsection{Detection of needle mispositioning}

\subsubsection{Needle shift}

\begin{figure}[htbp]
\begin{center}
\includegraphics[width=0.7\linewidth]{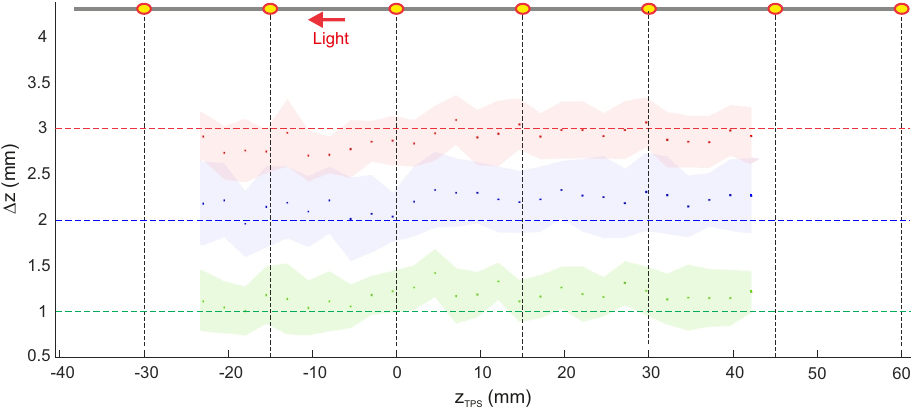}
\caption{Identification of source needle mispositioning: the deliberate needle shift, $\Delta z$, is plotted as a function of the planned dwell position along the z-axis ($z_{TPS}$). The predicted needle shifts of 1, 2, and 3 mm are represented by dashed lines, while the measured shifts are shown as square markers spaced at 2.5 mm intervals. The shaded areas indicate the standard deviation of the instantaneous source position across each 3.7-second dwell time of the treatment sequence. Vertical dashed lines indicate the probe positions, accompanied by a schematic of the detector at the top of the figure for reference, facilitating the localization of the analysis region relative to the 7SD. The treatment length and location remain consistent with the previous analyses.}\label{fig:shift}
\end{center}
\end{figure}

Figure \ref{fig:shift} illustrates the shift $\Delta z$ between the correctly positioned source needle and its deliberately mispositioned counterpart along the z-axis. On average, the intended needle shifts of 1, 2, and 3 mm were measured as 1.19 $\pm$ 0.27 mm, 2.20 $\pm$ 0.38 mm, and 2.90 $\pm$ 0.28 mm, respectively, across the evaluated source travel range.

\subsubsection{Needle tilt}

\begin{figure}[htbp]
\begin{center}
\includegraphics[width=0.7\linewidth]{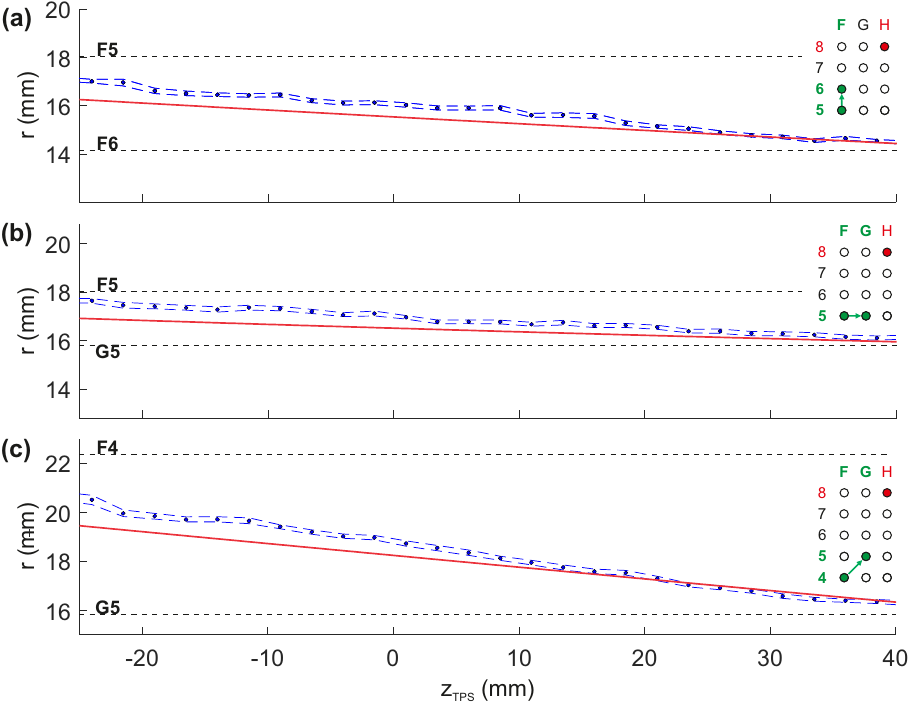}
\caption{Identification of source needle mispositioning: the deliberate needle tilt is analyzed by plotting the radial distance, $r$, of the source from the detector as a function of the planned dwell position, $z_{TPS}$, along the z-axis. The tilted source trajectory results from the tilted needle passing through the following hole pairs in the top and bottom templates: (a) F5 and F6, (b) F5 and G5, and (c) F4 and G5. Schematics of the corresponding tilt configurations within the templates are shown in the insets of each figure, with the 7SD consistently positioned in hole H8 of both the top and bottom templates. The horizontal black dashed lines indicate the distance between the probe hole (H8) and the two needle insertion holes, providing reference limits for the source travel range along $r$. Measured dwell positions are represented by blue points, while the blue dashed lines indicate measurement accuracy, quantified by the standard deviation of the instantaneous source position relative to the average dwell position. The red lines represent the theoretical straight path connecting the insertion holes through which the source needle passes.}\label{fig:tilt}
\end{center}
\end{figure}

The monitoring of a treatment sequence with a tilted source needle is presented in Fig. \ref{fig:tilt}. The three distinct needle tilt configurations are addressed in Figs.\ref{fig:tilt}(a), (b) and (c), respectively, and are depicted in the figure insets. The tilt analysis was conducted by measuring the source-probe distance, $r$, as a function of the planned altitude, $z_{TPS}$. The resulting functions, $r=f(z)$, exhibit a near-linear trend, with a little systematic curvature. The first dwell position of the treatment sequences is located at an altitude of $z = 39$ mm, which is 7 mm above the bottom solid-water template, cf. Figs. \ref{fig:exp2}(a) and (b). A systematic deviation is observed from the straight line connecting the two insertion holes in the top and bottom templates through which the source needle passes (shown in red in the figures). In all cases, as the source moves away from the bottom template, the measured points progressively deviate from the inter-hole connection lines, with the recorded $r$ values consistently exceeding those predicted by the theoretical straight path.

\newpage     

\section{discussion}

\subsection{Dwell identification}

Overall, 99.5\% of the source dwell positions were successfully identified, achieving a state-of-the-art dwell identification rate \cite{fonseca:piro20,johansen:pm19}. All dwell positions with dwell times greater than 0.2 seconds were detected, except for one. The missed dwell resulted from a time lag in the detection system, attributed to a Windows environment managing both camera control and data acquisition. To overcome lag issues, the camera and data recording processes should operate independently, each managed by a separate system, such as using a microcontroller or FPGA to control the camera, and a computer dedicated to handling data recording.

\subsection{Dwell position verification}

The slight increase in deviation from the planned source coordinates observed in Figs. \ref{fig:dwell_pos_xz}(a) and (c) for source altitudes below 7 mm can be attributed to a minor tilt of the source needle, the source tracking algorithm, or a slight misalignment (of 0.22 mm maximum) of the seven probes at the end of the 7-fiber bundle forming the 7SD. However, this variation in detection accuracy, which remains within 0.2 mm, is well bellow the reliability requirement of HDR-BT \cite{dempsey:recomm13}. Overall, the deviation from the planned dwell positions stays below 0.6 mm for both $x$ and $z$. 
Linares et al., using a similar triangulation approach, reported source position deviations of up to 1.6 mm along $z$ and 0.9 mm along $r$ over a 62 mm source displacement range, with dwell times limited to 10 seconds, albeit at a shorter source-probe spacing of 5 mm \cite{linares:mp20}. At such close proximity, higher fluctuations in source localization are expected due to the steep dose gradient near the source. In previous work, we demonstrated that a six-probe detector achieves comparable accuracy in dwell position determination at both 5.5 mm and 10 mm source-probe spacings for 1D source tracking along the $z$-axis with 10 s dwell times. Therefore one could expect similar performances of the 7SD at 5.5 mm and 10 mm from the source. 

The enhanced performance of our system is mainly due to its smaller scintillating detection volumes, between 110 and 280 times smaller than those used by Linares et al.\cite{linares:mp19,linares:mp20}, combined with a larger number of probes in our setup. This higher number of probes likely accounts for the consistent accuracy observed in the determination of the axial coordinate ($z$) (see Fig. \ref{fig:dwell_pos_xz}(c)). A slight decrease in the accuracy of radial coordinate ($r$) determination was observed as the source-probe spacing increased, in line with the anticipated reduction in SNR \cite{andersen:mp09_2}.

Expanding the number of probes is key to maintaining consistent accuracy in source position determination over source travel ranges extending beyond several centimeters. For instance, the three-point detector developed by Linares et al. showed source tracking accuracy that varied significantly with the source’s displacement along the detector at dwell times of 10 seconds \cite{linares:mp20}. In contrast, our previous results demonstrated that a six-probe detector eliminated this position dependence at identical dwell times \cite{gonod:mp23}, and the present study shows that such dependence remains negligible even for treatments involving dwell times as short as 0.1 seconds.

The tracking accuracy of the 7SD is closely influenced by the detector's SNR and the dwell time. Figures \ref{fig:dwell_pos_xz}(e,f), which assess the precision of the 7SD in tracking the instantaneous source position, demonstrate that the SNR is predominantly dependent on the source-probe spacing $r_{TPS}$. The noise level remained near constant throughout the entire range of the source's motion, irrespective of the values of $r_{TPS}$.

\subsection{Dwell time measurement}

On average, we observed similar performance to our previously developed single-probe detector, which shared the same probe geometry and detection architecture \cite{gonod:pmb22}. The main improvement in this study lies in the extension of the source-probe spacing $r_{TPS}$ from 20 mm, as previously reported, to 36 mm. Dwell time measurement relies on detecting signal edges between successive plateaus in the characteristic staircase-like temporal profile. This process is highly dependent on both the dose gradient and the inter-dwell spacing, which influence the sharpness and amplitude of these edges. As the source-probe spacing increases, both the dose gradient and the SNR decrease, leading to reduced edge detection capability and, consequently, less accurate dwell time measurements. The expanded monitoring range achieved here, using probes with the same individual performance, underlines the value of the 7SD’s multiprobe design. By leveraging seven independent probes simultaneously, the 7SD effectively multiplies the detection volume without compromising spatial resolution through volume averaging. This configuration allows for reliable dwell time verification over a broader range of source-probe spacings, reinforcing the potential of the detector for HDR-BT monitoring. 

Across the entire range of source-probe spacings from 10 to 36 mm, with dwell times varying between 0.1 s and 19.5 s and an inter-dwell spacing of 2.5 mm, we measured an average deviation from the planned dwell times of 0.006 $\pm$ 0.061 s. In comparison, Guiral et al. \cite{guiral:mp16} reported a deviation of 0.05 $\pm$ 0.9 s using a four-probe detector, with measurements performed at approximately 20 mm source-probe spacing, a detection interval of 0.1 s, 5-s dwell times, a source travel range of 60 mm, and an inter-dwell spacing of 2.5 mm. Additionally, Linares et al. \cite{linares:mp20} reported a deviation of 0.33 ± 0.37 s at a source-probe spacing of 5.5 mm, using 1-s dwell times and 1-mm inter-dwell spacing.

\subsection{Reliability}

\subsubsection{Systematic analysis} \label{sec:reliab}

In the context of the systematic analysis of the probes's reliability (Fig. \ref{fig:reliability}), the parameters $\Bar{R}$ and integration time serve as analogs for dwell positions and dwell times, respectively, offering a comprehensive evaluation of the system’s performance under clinically relevant conditions. We intentionally restricted the range of source-probe spacings to between 10 and 36 mm to cover the majority of prostate HDR-BT treatment volumes. In its current configuration, the 7SD detection system can potentially monitor treatments at larger source-probe spacings $r_{TPS}$, but with the limitation that only dwell times longer than 2 seconds can be reliably tracked. 

For source-probe distances beyond 36 mm, extending the camera acquisition time offers a straightforward way to compensate for reduced SNR and improve detector reliability. However, this adjustment limits the system's capacity to accurately detect shorter dwell times (e.g., under 0.5 seconds). An alternative solution could involve replacing the sCMOS camera with seven high-sensitivity detectors, such as photomultiplier tubes, to substantially boost SNR and expand the spatiotemporal  monitoring volume. Finally, the spatiotemporal monitoring volume could be significantly expanded by deploying at least two 7SDs, positioned several centimeters apart and fabricated from the same fiber bundle. The camera is particularly well-suited for this application, as it can simultaneously read signals from a large number of probes, regardless of their spatial arrangement. This enables the design of highly flexible multiprobe detection configurations without the need to modify the optical detection system itself. Moreover, such a setup would facilitate 3D source localization through triangulation techniques \cite{linares:mp21}. Therefore, viable strategies are available to achieve optimal detection reliability across the treatment volumes and the full spectrum of dwell times used in most HDR-BT procedures. 

\subsubsection{Treatment plan} 

The subtle color gradients observed in Fig. \ref{fig:rel_treat}(b) and (c) suggest only a minor trend of increasing deviation from the planned dwell positions with larger source-probe spacings. This trend may be attributed to a slight reduction in tracking accuracy primarily affecting the $r$ (radial) coordinate of the source (cf. Fig. \ref{fig:dwell_pos_xz}(d)).

Under treatment plan conditions, 99.4\% of the identified dwells were located with an error below the 1-mm reliability threshold, even with dwell times as short as 0.1 s (Fig. \ref{fig:rel_treat}(a)). Surprisingly, the largest dwell position offsets were not observed at the shortest dwell times but rather within the dwell time range between 1 s to 15 s (Fig. \ref{fig:rel_treat}(b)). Fig. \ref{fig:rel_treat}(c) shows that, with a single exception, all deviations exceeding the 1-mm reliability threshold occurred at the initial dwell position (dwell $\#1$) of the treatment sequences. Specifically, 30.4\% and 8.9\% of the dwell $\#1$ positions displayed offsets exceeding 0.77 mm and 1 mm, respectively. These rates drop to 1.4\% and 0.35\% for the second dwell positions (dwell $\#2$), with no further dwell positions surpassing the 0.77 mm threshold. The clustering of data points exceeding 0.77 mm for dwell $\#1$ suggests a potential positioning error of the afterloader specifically occurring at the initial dwell position of the treatment sequences. Further investigation is warranted to determine whether this afterloader positioning error is systematic. Excluding these critical dwell positions, the 7SD achieved 100\% reliability across the five prostate HDR-BT treatment plans assessed in this study.

\subsection{Detection of needle mispositioning}

\subsubsection{Needle shift}

The discrepancy between the planned and measured needle shifts remained within the measurement uncertainty of the 7SD (indicated by the standard deviation shown in the shaded area of Fig. \ref{fig:shift}) and the positional accuracy of the afterloader \cite{robinson:brachy21}. These findings confirm that the 7SD can reliably detect and monitor needle shifts on the millimeter scale.

\subsubsection{Needle tilt}

The systematic position-dependent deviation from the straight line connecting the two insertion holes, which diminishes as the dwell positions approach the bottom template, reflects the S-shaped bending of the needle caused by the insertion misalignment. The 7SD demonstrates sufficient accuracy to clearly detect sub-millimeter amplitude bending along the interstitial needle.

\subsection{Uncertainty budget}

See Table \ref{tab:budget}.
The expanded uncertainty ($k=3$), corresponding to a 99.7\% confidence level within a normal probability distribution, remains around 1 mm, closely matching the reliability performance of the 7SD, where 100\% of the dwell positions were measured with an offset of less than 1 mm from the planned dwell positions (cf. Fig. \ref{fig:rel_treat}(a)).


\begin{table}
\footnotesize
\begin{center}
\begin{tabular}{c c c c c c}


\hline
\multicolumn{2}{c}{ }  & Dwell position &    Dwell time   &     & \\

\multicolumn{2}{c}{ }   & uncertainty   &   uncertainty  & Type &  Source   \\

\multicolumn{2}{c}{ }  &     (mm)       &      (s)       &      &  \\

\hline
\textbf{Detector} & Repeatability &      &       &      &  \\
 
                  &       (0z)       & 0.182&  --     &   A   &  Fig. \ref{fig:dwell_pos_xz}  \\
                  &       (0r)       & 0.155&   --    &   A   &  Fig. \ref{fig:dwell_pos_xz} \\
                  &      --          &  --  &  0.061  &   A   &  Fig. \ref{fig:dwell_time} \\

                  & Scint. temp.    &       &         &        &   \\
               & (17-19$^{\circ}$C) &       &         &        &  \\
 
               &        (0z)         & 0.015      &  --     & B      &  Ref. \cite{mclaughlin:rm23}, Sec. \ref{sec:calib} \\
 
               &        (0r)         & 0.010       & --    & B       & Ref. \cite{mclaughlin:rm23}, Sec. \ref{sec:calib} \\
 
               & Stem effect         & 0         & 0      & B         & Sec. \ref{sec:readout} \\

               & Integration         & --        & 0.017     & B        &  Sec. \ref{sec:readout} \\ 
               & time                &           &(0--0.06)&            &     \\ 
               & \textbf{Total}      & \textbf{0.240} & \textbf{0.063} &    &  \\
 
\hline

\textbf{Needle/detector}  & Needle/(0r) & 0.029        & -- & B & Sec. \ref{sec:encaps}  \\
\textbf{mispositionning } &             & ($\pm$ 0.05) &    &   &   \\

                          & Detector/(0r) & 0.104         & -- & B &  Sec. \ref{sec:encaps} \\
                          &               & ($\pm$ 0.18) &    &   &   \\

                          & \textbf{Total} &\textbf{0.108} & -- &  &  \\
\hline
 
\textbf{Afterloader} & Source        &  0.231       &  0.058         &    B   &  Ref. \cite{robinson:brachy21} \\

                     & position/(0z) & ($\pm$0.4)   &  ($\pm$0.1)    &        &    \\

 \hline\hline

\textbf{Overall} (k=1 -- 68.3\%)    &               & \textbf{0.350} & \textbf{0.086} &  &  \\
\textbf{Overall} (k=3 -- 99.7\%)    &               & \textbf{1.050} & \textbf{0.258} &  &  \\
\hline
\end{tabular}
\end{center}
\caption{Uncertainty budget for dwell time and position verification, as presented by Andersen et al. \cite{andersen:mp09}. In accordance with ISO (1995) guidelines, uncertainties are classified into two categories: Type A, which are evaluated using valid statistical methods, and Type B, which are based on scientific judgment and available information. A Type B uncertainty, $u$, is calculated as $u = (v_{\text{max}} - v_{\text{min}})/\sqrt{12}$, where $v_{\text{max}}$ and $v_{\text{min}}$ represent the maximum and minimum values, respectively. The total combined uncertainty, $u_{\text{tot}}$, for a series of uncertainties ($u_i$, where $i \in N$), is determined by $u_{\text{tot}} = \sqrt{\sum_i u_i^2}$.}\label{tab:budget}
\end{table}

\subsection{Clinical use}

Having passed cytotoxicity testing and confirmed as biocompatible, the 7SD is well-suited for clinical use in HDR brachytherapy. Its compact dimensions allowed insertion into a biocompatible plastic sheath narrower than standard interstitial needles, making it compatible with most HDR-BT needles and catheters, including integration into clinical applicators.

Further studies are needed to evaluate the probe's performance under a broader range of clinical conditions, particularly when using sources approaching the end of their clinical lifespan (around 3 to 4 months). In such scenarios, reduced source activity results in lower SNR and necessitates longer dwell times to deliver the prescribed dose. These additional investigations will help validate the 7SD’s reliability and effectiveness in lower-activity treatments.

We established a stringent reliability threshold for the 7SD, requiring that 99.7\% of the dwell position deviations stay within 1 mm (equivalent to $3\sigma$ of a normal distribution), to anticipate additional uncertainties in clinical HDR-BT monitoring that may alter the probe. While our study used a single-needle setup, clinical prostate brachytherapy typically involves multiple manually-implanted needles, which could alter the local energy distribution and attenuate the 7SD signal. Moreover, the accuracy of in vivo HDR-BT monitoring is highly dependent on the positional stability of the 7SD, which could be affected by anatomical shifts during the treatment \cite{fonseca:piro20}. This limitation would likely be alleviated in applicator-based brachytherapy, where both the source and detector catheters would be securely fixed relative to each other, reducing the potential for positional drifts. Additionally, triangulation algorithms could be refined using X-ray, MRI, or ultrasound imaging, which are routinely used to visualize needles or catheters \textit{in vivo}, either in real time or during pre- and post-treatment assessments.

In its current configuration, the 7SD offers a monitoring volume sufficient for most HDR-BT clinical applications. However, its lateral detection range may be limited in cervical cancer treatments requiring complex applicators with laterally positioned implantation needles. Several strategies described in Section \ref{sec:reliab} could be employed to broaden the detection volume for these challenging geometries. While the 7SD’s monitoring field fully covers the prostate HDR-BT treatment volume  when placed at the center of the BT template (cf. Fig. \ref{fig:exp2}(c)), this approach may not be clinically feasible for the smaller prostates, where all central template holes are required for treatment needles, leaving no available space for detector placement \cite{mason:ro16}. Therefore, as outlined in Section \ref{sec:reliab}, more advanced configurations incorporating at least two 7SD detectors placed at the periphery of the treatment volume could be used to perform remote monitoring throughout the tumor region. This setup would also enable 3D source localization via triangulation techniques \cite{linares:mp21}. 

The proposed multiprobe detection strategy provides outstanding flexibility and scalability, enabling the development of optimized detection systems for a wide variety of HDR-BT techniques. By adjusting parameters such as inter-probe spacing, the number of probes per detector, and the overall number of detectors, the system can be tailored to meet specific clinical requirements. Probes may be integrated into a single fiber bundle, multiple bundles, or distributed separately. The camera-based readout system maximizes modularity and scalability by supporting a wide range of detection configurations -- with varying probe numbers and architectures -- using a single optical sensor.

\section {Conclusion}

We presented the performance of a 7SD system for time-resolved HDR-BT monitoring across a broad range of dwell times (0.1 to 19.5 seconds), a 62 mm source travel distance, and source-detector spacings up to 36 mm. Compared to single-probe detectors \cite{gonod:pmb22}, the 7SD’s multi-probe design significantly expands the monitoring volume both axially and radially and enables accurate 2D source tracking through triangulation, defining state-of-the-art performance levels. The detector’s biocompatibility was confirmed through cytotoxicity testing. Of the 4040 dwell positions analyzed in this study, 99.5\% were successfully identified, with a 100\% identification rate for dwell times exceeding 0.2 s. On average, dwell position determination achieved an accuracy of 0.224 $\pm$ 0.155 mm along the source displacement axis and 0.077 $\pm$ 0.181 mm in the orthogonal radial direction. The mean deviation from planned dwell times was 0.006 $\pm$ 0.061 s. A high reliability level was demonstrated, with 99.4\% of all dwell positions measured within the 1 mm threshold. The remaining 0.6\% of deviations, consistently observed at the initial dwell positions of treatment sequences, appear to result from a systematic source positioning error by the afterloader. The consistent accuracy of the 7SD across the entire source travel range enables the detection of subtle afterloader malfunctions, providing a precise and valuable tool for assessing the integrity of HDR-BT clinical setups and procedures. Additionally, the detector precisely identified intentional needle mispositioning scenarios, including both shifts and tilts, with submillimeter accuracy. This study demonstrates that the proposed 7SD system is well-suited for use as an \textit{in vivo} detector for time-resolved HDR-BT treatment monitoring. It delivers state-of-the-art performance while offering exceptional flexibility and scalability to accommodate a wide range of HDR-BT treatment scenarios.

\clearpage

\section*{Acknowledgments}
This work has received funding from the SAYENS Agency, the French Agence Nationale de la Recherche under project  NANOPTiX (ANR-18-CE42-0016), the EQUIPEX+ SMARTLIGHT platform (ANR-21-ESRE-0040), the EQUIPEX+ NANOFUTUR (ANR-21-ESRE-0012) and the EIPHI Graduate School (ANR-17-EURE-0002). This work was also supported by the French Renatech network, MIMENTO technological facility, and the Région Bourgogne Franche-Comté.

\section*{References}






\end{document}